\documentclass[aps,pra,twocolumn,showpacs,groupedaddress]{revtex4-1}
\usepackage{amsfonts}
\usepackage{amssymb}
\usepackage{amsmath}
\usepackage{graphicx}
\usepackage{xcolor}

\begin{document}

\title{Analytical solutions of the coupled Gross-Pitaevskii equations for
the three-species Bose-Einstein condensates}

\author{Y.M.Liu$^{1,3}$ and C.G.Bao$^{2}$}
\thanks{Corresponding author: C.G.Bao, stsbcg@mail.sysu.edu.cn}

\affiliation{$^1$Department of Physics, Shaoguan University, Shaoguan, 512005, P. R. China}
\affiliation{$^2$State Key Laboratory of Optoelectronic Materials and Technologies, School of
Physics and Engineering, Sun Yat-Sen University, Guangzhou, P. R. China}
\affiliation{$^3$State Key Laboratory of Theoretical Physics, Institute of Theoretical
Physics, Chinese Academy of Sciences, Beijing, 100190, China}


\begin{abstract}
The coupled Gross-Pitaevskii equations for the g.s. of the three-species
condensates (3-BEC) have been solved analytically under the Thomas-Fermi
approximation. Six types of spatial configurations in miscible phase are found.
The whole parameter-space has been divided into
zones each supports a specific configuration (miscible or immiscible).
The borders of the zones are described
by analytical formulae. Due to the division, the variation of the spatial
configuration against the parameters can be visualized, and the effects of the parameters can be thereby
understood. There are regions in the parameter-space where the configuration
is highly sensitive to the parameters. These regions are tunable and valuable for the determination of the parameters.
\end{abstract}


\maketitle

\section{Introduction}

In recent years there are a number of literatures dedicated to the
theoretical \cite{ho96,esry97,pu98,luo07,nott15,
scha15,inde15,kuop15,roy15,luo08,polo15}. and experimental \cite%
{myat97,ande05,ni08,pilc09,nemi09,wack15} study of the two-species
Bose-Einstein condensates (2-BEC) (also refer to the references in \cite%
{wack15}). The 2-BEC provides an important tool to clarify the inter-species
and intra-species interactions. The miscible and immiscible
phases of the ground state (g.s.) have been predicted and have been experimentally
confirmed \cite{ni08}. On the other hand, the study on the condensates with
more than two species is very scarce.\cite{cal,man} Since the multi-species
BEC is in principle experimentally achievable \cite{tag}, a primary
theoretical attempt might be worthy to see whether interesting physics is
involved and whether this new field deserves a further study. This paper is
dedicated to this purpose, namely, a primary theoretical study on the
three-species Bose-Einstein condensates (3-BEC).

The spatial configurations of the 3-BEC, as in 2-BEC, are expected to have also three phases:
 miscible, immiscible, and asymmetric phases (as shown below). For the
first phase the atoms of each kind of species are compactly distributed
surrounding the center of the trap, the distribution of some species is
broader and some narrower. For the second either at least one species leaves
completely from the center or at least one species is distributed in more
than one disconnected spatial\ domains. For the first and second, the
distribution keeps the symmetry as the trap. For the third, the distribution
does not keep the symmetry of the trap. This paper is dedicated to the g.s. in
miscible phase. The emphasis is placed on the qualitative aspect. The trap is
assume to be isotropic. The spin-degrees of freedom are frozen. By
introducing the Thomas-Fermi approximation (TFA, in which the kinetic energy
is neglected), the coupled Gross-Pitaevskii equations (CGP) for the g.s.
 are solved analytically. This enable us to carry on the
analysis in an analytical way. According to the relative distributions of
the three species, the miscible phase has been further classified into six
types. The analytical formalism derived in the follows enable us to divide
the whole parameter-space into zones, each supports a specific type. Based
on the division, the variety of the spatial configurations and their variation against the
parameters (the intra- and inter-species interactions, the particle numbers,
masses, and those for the trap) can be visualized, and thereby the effect of these
 parameters can be clarified.

\section{Hamiltonian and the coupled Gross-pitaevskii equations}
We consider three kinds of atoms $N_{A}$ A-atoms with mass $m_{A}$\ and
interacting via $V_{A}=c_{A}\Sigma _{i<i^{\prime }}\delta (\mathbf{r}_{i}%
\mathbf{-r}_{i^{\prime }})$, $N_{B}$ B-atoms with $m_{B}$, $V_{B}$, and $%
c_{B}$, and $N_{C}$ C-atoms with $m_{C}$, $V_{C}$, and $c_{C}$. The particle
numbers are assumed to be huge (say, larger than 10000). The interspecies
interactions are $V_{AB}=c_{AB}\Sigma _{i<j}\delta (\mathbf{r}_{i}\mathbf{-r}%
_{j})$ with the strength $c_{AB}$, $V_{BC}$ with $c_{BC}$, and $V_{CA}$ with $%
c_{CA}$. These atoms are confined by the harmonic traps $\frac{1}{2}%
m_{s}\omega _{s}^{2}r^{2}$ ( $s=A$, $B$ or $C$). We introduce a mass $m_{o}$
and a frequency $\omega $. Then, $\hbar \omega $ and $\lambda \equiv \sqrt{%
\hbar /(m_{o}\omega )}$ are used as units for energy and length in this
paper. The total Hamiltonian is

\begin{eqnarray}
H=H_{A}+H_{B}+H_{C}+V_{AB}+V_{BC}+V_{CA}  \nonumber \\
H_{A}=\sum_{i=1}^{N_{A}}(-\frac{m_{o}}{2m_{A}}\nabla _{i}^{2}+\frac{1}{2}%
\gamma _{A}r_{i}^{2})+V_{A}  \label{eq1}
\end{eqnarray}
where $\gamma _{A}=(m_{A}/m_{o})(\omega _{A}/\omega )^{2}$. $H_{B}$ and $%
H_{C}$ are similarly defined.

We consider the g.s. in which no spatial excitations are
involved. Thus, each kind of atoms are fully condensed into a state which is
most advantageous for binding (otherwise, the energy would be higher).
Accordingly, the total wave function of the g.s. can be written as

\begin{equation}
\Psi =\Pi _{i=1}^{N_{A}}\frac{u_{1}(r_{i})}{\sqrt{4\pi }r_{i}}\Pi
_{j=1}^{N_{B}}\frac{u_{2}(r_{j})}{\sqrt{4\pi }r_{j}}\Pi _{k=1}^{N_{C}}\frac{%
u_{3}(r_{k})}{\sqrt{4\pi }r_{k}}  \label{eq2}
\end{equation}
where $u_{1}$, $u_{2}$, and $u_{3}$ are for the A-, B-, and
C-atoms, respectively.

From minimizing the total energy, we obtain the set of CGP. One of them is

\begin{eqnarray}
(-\frac{m_{o}}{2m_{A}}\nabla ^{2}+\frac{1}{2}\gamma _{A}r^{2}+N_{A}c_{A}%
\frac{u_{1}^{2}}{4\pi r^{2}}+N_{B}c_{AB}\frac{u_{2}^{2}}{4\pi r^{2}} \nonumber \\
+N_{C}c_{CA}\frac{u_{3}^{2}}{4\pi r^{2}}-\varepsilon _{A})u_{1}=0
\label{eq3}
\end{eqnarray}
where $\varepsilon _{A}$\ is the chemical potential. Via cyclic permutations
of the three indexes $(A,B,C)$ and the three $(u_{1},u_{2},u_{3})$, from eq.(%
\ref{eq3}) we obtain the other two equations. It is emphasized that the three equations of
normalization ${\int u_{l}^{2}dr=1}$ ($l$=1, 2, and 3) should hold.

\section{Formal solutions under the Thomas-Fermi approximation}

Since $N_{A}$, $N_{B}$ and $N_{C}$ are considered to be large, the
approximation TFA has been adopted. A recent numerical evaluation of this
approximation is referred to the papers \cite{yzhe,polo15}. Under the TFA,
the CGP become

\begin{eqnarray}
(\frac{r^{2}}{2}+\alpha _{11}\frac{u_{1}^{2}}{r^{2}}+\alpha _{12}\frac{%
u_{2}^{2}}{r^{2}}+\alpha _{13}\frac{u_{3}^{2}}{r^{2}}-\varepsilon
_{1})u_{1}=0  \nonumber\\
(\frac{r^{2}}{2}+\alpha _{21}\frac{u_{1}^{2}}{r^{2}}+\alpha _{22}\frac{%
u_{2}^{2}}{r^{2}}+\alpha _{23}\frac{u_{3}^{2}}{r^{2}}-\varepsilon
_{2})u_{2}=0 \nonumber \\
(\frac{r^{2}}{2}+\alpha _{31}\frac{u_{1}^{2}}{r^{2}}+\alpha _{32}\frac{%
u_{2}^{2}}{r^{2}}+\alpha _{33}\frac{u_{3}^{2}}{r^{2}}-\varepsilon
_{3})u_{3}=0  \label{eq4}
\end{eqnarray}
where $\alpha _{11}=N_{A}c_{A}/(4\pi \gamma _{A})$, $\alpha
_{22}=N_{B}c_{B}/(4\pi \gamma _{B})$, $\alpha _{33}=N_{C}c_{C}/(4\pi \gamma
_{C})$, $\alpha _{12}=N_{B}c_{AB}/(4\pi \gamma _{A})$, $\alpha
_{21}=N_{A}c_{AB}/(4\pi \gamma _{B})$, $\alpha _{13}=N_{C}c_{CA}/(4\pi
\gamma _{A})$, $\alpha _{31}=N_{A}c_{CA}/(4\pi \gamma _{C})$, $\alpha
_{23}=N_{C}c_{BC}/(4\pi \gamma _{B})$, $\alpha _{32}=N_{B}c_{BC}/(4\pi
\gamma _{C})$, they are called the weighted strengths (W-strengths). $%
\varepsilon _{1}=\varepsilon _{A}/\gamma _{A}$, $\varepsilon
_{2}=\varepsilon _{B}/\gamma _{B}$, $\varepsilon _{3}=\varepsilon
_{C}/\gamma _{C}$, they are the weighted energies for a single particle. In
this paper all the interactions are considered as repulsive. Accordingly,
all the W-strengths are positive. Furthermore, it is safe to assume that all
the $u_{l}/r$ are always non-negative. Recall that there are originally 15
parameters ($N_{s},m_{s},\omega _{s},c_{s},c_{ss^{\prime }}$). From eq.(%
\ref{eq4}) we know that their combined effects are fully
represented by the nine $\alpha _{ll^{\prime }}$. Among them, only eight are
independent because they are related as $\alpha _{12}\alpha _{23}\alpha
_{31}=\alpha _{21}\alpha _{32}\alpha _{13}$. Thus, based on the W-strengths, related analysis could be
simpler.

We define a matrix $\mathfrak{M}$ with its element $(\mathfrak{M}%
)_{ll^{\prime }}=\alpha _{ll^{\prime }}$. The determinant of $\mathfrak{M}$
is denoted by $\mathfrak{D}$. The algebraic cominor of $\alpha _{ll^{\prime
}}$ is denoted as $d_{ll^{\prime }}$. Obviously, the element of the inverse
matrix $(\mathfrak{M}^{-1})_{ll^{\prime }}=d_{l^{\prime }l}/\mathfrak{D}$.

The set of equations (\ref{eq4}) has four forms of formal solution, each
holds in a specific domain of $r$:

(i) Form III: When all the three wave functions are nonzero in a domain, they
must have the form as

\begin{equation}
u_{l}^{2}/r^{2}=X_{l}-Y_{l}r^{2}  \label{eq5}
\end{equation}
where
\begin{equation}
X_{l}=\mathfrak{D}_{X_{l}}\mathfrak{/D}  \label{eq6}
\end{equation}
$\mathfrak{D}_{X_{l}}$ is a determinant obtained by changing the $l$\ column
of $\mathfrak{D}$ from $(\alpha _{1l},\alpha _{2l},\alpha _{3l})$ to $%
(\varepsilon _{1},\varepsilon _{2},\varepsilon _{3})$.
\begin{equation}
Y_{l}=\mathfrak{D}_{Y_{l}}\mathfrak{/D}  \label{eq7}
\end{equation}
$\mathfrak{D}_{Y_{l}}$ is also a determinant obtained by changing the $l$\
column of $\mathfrak{D}$ to $(1/2,1/2,1/2)$. Once all the parameters are
given, the three $Y_{l}$ are known because they depend only on $\alpha
_{ll^{\prime }}$. However, the three ${X_{l}}$ have not yet been known because they depend also on
$\varepsilon _{1}$ to $\varepsilon _{3}$. When $Y_{l}$\ is positive (negative), $u_{l}/r$\ goes down
(up) with $r$. Obviously, once $Y_{l}$\ is positive, $X_{l}$ must be large
enough to prevent $u_{l}/r$\ to be negative.

(ii) Form II: Let $(l,m,n)$\ be a cyclic permutation of (1,2,3), the same in
the follows. When one and only one of the wave functions is zero inside the domain (say, $%
u_{n}/r=0$), the other two must have the form as

\begin{eqnarray}
u_{l}^{2}/r^{2} &=&X_{l}^{(n)}-Y_{l}^{(n)}r^{2}  \nonumber\\
u_{m}^{2}/r^{2} &=&X_{m}^{(n)}-Y_{m}^{(n)}r^{2}  \label{eq8}
\end{eqnarray}
where
\begin{eqnarray}
X_{l}^{(n)} &=&(\alpha _{mm}\varepsilon _{l}-\alpha _{lm}\varepsilon
_{m})/d_{nn}  \nonumber\\
Y_{l}^{(n)} &=&\frac{1}{2}(\alpha _{mm}-\alpha _{lm})/d_{nn}  \nonumber\\
X_{m}^{(n)} &=&(\alpha _{ll}\varepsilon _{m}-\alpha _{ml}\varepsilon
_{l})/d_{nn}  \nonumber\\
Y_{m}^{(n)} &=&\frac{1}{2}(\alpha _{ll}-\alpha _{ml})/d_{nn}  \label{eq9}
\end{eqnarray}

Once the parameters are given, the six $Y_{n^{\prime }}^{(n)}$ ($n^{\prime
}\neq n$) are known because they depend only on $\alpha _{ll^{\prime }}$.
When $Y_{n^{\prime }}^{(n)}$ is positive (negative), $u_{n^{\prime }}/r$\
goes down (up) with $r$. Obviously, once $Y_{n^{\prime }}^{(n)}$\ is
positive, the unknowns $X_{n^{\prime }}^{(n)}$ must be positive and large enough.

(iii) Form I: When\ one and only one of the wave functions is nonzero in a
domain (say, $u_{l}/r\neq 0$), it must have the form as

\begin{equation}
u_{l}^{2}/r^{2}=\frac{1}{\alpha _{ll}}(\varepsilon _{l}-r^{2}/2)
\label{eq10}
\end{equation}

Obviously, $u_{l}/r$ in this form must descend with $r$. This form could
emerge only if $\varepsilon _{l}$ is positive and sufficiently large.

(iv) Form 0: In this form\ all the three wave functions are zero.

If $u_{l}/r$\ is nonzero in a domain but becomes zero when $r=r_{o}$, then a
downward form-transition (say, from Form III to II) will occur at $r_{o}$.
Whereas if $u_{l}/r$\ is zero in a domain but becomes nonzero when $r=r_{o}$%
, then a upward form-transition (say, from Form II to III) will occur at $%
r_{o}$. In this way the formal solutions will link up continuously to form
an entire solution. They are continuous at the transition points because the
wave functions satisfy exactly the same set of
nonlinear equations at $r_{o}$. However, their derivatives are in general not
continuous at $r_{o}$.

When all the W-strengths are given, however, there are three unknowns $%
{\varepsilon _{l}}$ contained in the entire solution. Once they are known all
the ${X_{l}}$ and ${X_{l^{\prime }}^{(l)}}$ can also be known. Due to the
requirement of normalization, we have three additional equations. They are
sufficient to determine the three ${\varepsilon _{l}}$ as shown below.

\section{Three lemmas}

There are three lemmas related to the linking of formal solutions.

\textit{Lemma I: The three }${Y_{l}}$ \textit{can not all be negative.}

Let us define a vector $\overset{\rightarrow }{\Omega _{l}}\equiv \alpha
_{1l}\overset{\rightarrow }{n_{1}}+\alpha _{2l}\overset{\rightarrow }{n_{2}}%
+\alpha _{3l}\overset{\rightarrow }{n_{3}}$, where $(\overset{\rightarrow }{%
n_{1}},\overset{\rightarrow }{n_{2}},\overset{\rightarrow }{n_{3}})$ are a
set of orthogonal unit vectors, and all the $\alpha _{ll^{\prime }}$ are
assumed to be positive as mentioned. Therefore $\overset{\rightarrow }{%
\Omega _{l}}$ is situated inside the first octant. It can be rewritten as $%
\overset{\rightarrow }{\Omega _{l}}=|\Omega _{l}|\overset{\rightarrow }{q_{l}%
}$, where $\overset{\rightarrow }{q_{l}}$ is also a unit vector in the first
octant. We define further $\overset{\rightarrow }{n}\equiv \frac{1}{2}(%
\overset{\rightarrow }{n_{1}}+\overset{\rightarrow }{n_{2}}+\overset{%
\rightarrow }{n_{3}})$. Then, $Y_{l}=\frac{\overset{\rightarrow }{n}\cdot (%
\overset{\rightarrow }{q_{m}}\times \overset{\rightarrow }{q_{n}})}{|\Omega
_{l}|\overset{\rightarrow }{q_{l}}\cdot (\overset{\rightarrow }{q_{m}}\times
\overset{\rightarrow }{q_{n}})}$. The three $\overset{\rightarrow }{q_{l}}$,
$\overset{\rightarrow }{q_{m}}$, and $\overset{\rightarrow }{q_{n}}$ should
be linearly independent (otherwise, the determinant $\mathfrak{D}$ is zero
and the Form III does not exist). Then, $\overset{\rightarrow }{n}$ can be
expanded as
\begin{equation*}
\overset{\rightarrow }{n}=n_{l}\overset{\rightarrow }{q_{l}}+n_{m}\overset{%
\rightarrow }{q_{m}}+n_{n}\overset{\rightarrow }{q_{n}}  \label{nl}
\end{equation*}%
and accordingly
\begin{equation*}
Y_{l}=\frac{n_{l}}{|\Omega _{l}|}  \label{yl}
\end{equation*}
Thus, the sign of $Y_{l}$\ is determined by $n_{l}$.

Since all the three $\overset{\rightarrow }{q_{l}}$ to $\overset{\rightarrow
}{q_{n}}$ are inside the first octant, if all the three $n_{l}$, $n_{m}$,
and $n_{n}$ were negative, $-\overset{\rightarrow }{n}$ would be in the
first octant. This is in contradiction with the definition of $\overset{%
\rightarrow }{n}$. Thus the three ${{Y_{l}}}$ can not all be negative, and the
lemma is proved.

This lemma implies that Form III must transform to Form II somewhere because at least
one of the ${Y_{l}}$ is positive, and therefore at least one the wave functions is descending
 and eventually arrives at zero.

\textit{Lemma II: }$Y_{m}^{(l)}$ \textit{and }$Y_{n}^{(l)}$ \textit{can
not both be negative.}

When $l=3$, we define three 2-dimensional vectors $\overset{\rightarrow }{%
\omega }_{t}\equiv \alpha _{1t}\overset{\rightarrow }{n_{1}}+\alpha _{2t}%
\overset{\rightarrow }{n_{2}}$ ($t=1$, 2) and $\overset{\rightarrow }{n_{12}}%
\equiv \frac{1}{2}(\overset{\rightarrow }{n_{1}}+\overset{\rightarrow }{n_{2}%
})$. All of them are situated in the first quadrant. Then, $Y_{1}^{(3)}=%
\frac{\overset{\rightarrow }{n_{3}}\cdot (\overset{\rightarrow }{n_{12}}%
\times \overset{\rightarrow }{\omega _{2}})}{\overset{\rightarrow }{n_{3}}%
\cdot (\overset{\rightarrow }{\omega _{1}}\times \overset{\rightarrow }{%
\omega _{2}})}$ and $Y_{2}^{(3)}=\frac{\overset{\rightarrow }{n_{3}}\cdot (%
\overset{\rightarrow }{\omega _{1}}\times \overset{\rightarrow }{n_{12}})}{%
\overset{\rightarrow }{n_{3}}\cdot (\overset{\rightarrow }{\omega _{1}}%
\times \overset{\rightarrow }{\omega _{2}})}$. $Y_{1}^{(3)}<0$ implies that,
on the $\overset{\rightarrow }{n_{1}}\overset{\rightarrow }{-n_{2}}$ plane,
the polar angle of $\overset{\rightarrow }{\omega }_{2}$ should lie between
those of $\overset{\rightarrow }{n_{12}}$ and $\overset{\rightarrow }{\omega
}_{1}$. Whereas $Y_{2}^{(3)}<0$ implies that the polar angle of $\overset{%
\rightarrow }{\omega }_{1}$ should lie between those of $\overset{%
\rightarrow }{n_{12}}$ and $\overset{\rightarrow }{\omega }_{2}$. These two
requirements are in contradiction. The cases with $l\neq 3$ are similar.
Thus the lemma is proved. In fact, this lemma can also be directly proved
via the definition of $Y_{m}^{(l)}$\ and $Y_{n}^{(l)}$.

This lemma implies that Form II will transform to Form I somewhere because at least one of the
wave functions is descending. Otherwise, it will transform to Form III if the missing wave function
emerges.
This lemma implies that Form II will either transform to Form I somewhere because at least one of the
wave functions (say, $u_{n}/r$) is descending, or transform to Form III if the missing wave function
emerges earlier than the vanish of $u_{n}/r$.

\textit{Lemma III: In a domain (or at a point) where all the three }${u_{l}/r}$%
 \textit{are zero, no wave function can emerge and becomes nonzero in this
domain (at the point).}

If $u_{l}/r$ emerges singly, then it must have the form eq.(\ref{eq10}),
therefore $u_{l}/r$ must descend with $r$\ and the emergence fails. If $u_{l}/r$
and $u_{m}/r$\ emerge in pair at the same place, then both $Y_{l}^{(n)}$\
and $Y_{m}^{(n)}$ should be negative to assure the uprising. This fails due
to \textit{Lemma II}. If all the three ${u_{l}/r}$ emerge together at the
same place, then all the three ${Y_{l}}$ should be negative to assure the
uprising. This fails due to \textit{Lemma I}. Thus, the \textit{Lemma III}
is proved.

Due to \textit{Lemma III}, once the unique nonzero wave function in Form I
arrives at zero, say, $u_{l}/r=0$\ when $r=r_{out}$, then $r_{out}$ will be
the outmost border for all kinds of atoms.

\section{ Linking the formal solutions to form an entire solution in miscible phase}

With the three lemmas, we are going to link up the formal solutions to form an
entire solution. To this aim, we will first make some presumptions so that the formal
solutions can be linked up in a specific way.
Then, we find out a subspace in the whole parameter-space. When the parameters are chosen inside this
  subspace, all the presumptions can be recovered so that the entire solution stands.
 In this way the whole space is divided into zones each supports a specific
spatial configuration of the g.s.. Based on the division, we are able to obtain various types of phase-diagrams
 to demonstrate the variation of the g.s. against
the parameters.

For the miscible phase, the first domain (starting from $r=0$) must
have Form III. Therefore, the three ${X_{l}}>0$ should be presumed. Due to
\textit{Lemma I}, there is at least a positive $Y_{l}$. Without loss of
generality, it is assigned that $X_{l}/Y_{l}$ is the smallest positive ratio
among the three ratios. Accordingly, among the three wave functions, $%
u_{l}/r $\ will arrive at zero first (refer to eq.(\ref{eq5})). Thus, the
first domain ends at $r_{a}\equiv \sqrt{X_{l}/Y_{l}}$, where a downward
form-transition occurs. For miscible phase $u_{l}/r$ is not allowed to
emerge again because it is not allowed to distribute in disconnected region.
Therefore $u_{l}/r$ is distributed only in $(0,r_{a})$. From the
normalization $\int_{0}^{r_{a}}u_{l}^{2}dr=%
\int_{0}^{r_{a}}(X_{l}r^{2}-Y_{l}r^{4})dr=1$, we have $%
X_{l}=(15/2)^{2/5}Y_{l}^{3/5}$ and
\begin{equation}
r_{a}^{2}=(\frac{15}{2Y_{l}})^{2/5}  \label{eq11}
\end{equation}

\noindent This equation implies that the W-strengths should be so chosen that $%
Y_{l}\geq Y_{m}$ and $Y_{l}\geq Y_{n}$ hold. This choice assures that $%
u_{l}/r$\ will arrive at zero first and the presumption $X_{l}>0$ can be
recovered.

The second domain will have the Form II and starts from $r_{a}$. Since at
least one of the two wave functions must descend with $r$ (\textit{Lemma II}%
), we can assign the one that arrives at zero first with the index $m$, and
we define $r_{b}\equiv \sqrt{X_{m}^{(l)}/Y_{m}^{(l)}}$ (refer to eq.(%
\ref{eq8})). Then, the equation of normalization for
$u_{m}$ is%
\begin{equation*}
1=\int_{0}^{r_{a}}(X_{m}r^{2}-Y_{m}r^{4})dr+%
\int_{r_{a}}^{r_{b}}(X_{m}^{(l)}r^{2}-Y_{m}^{(l)}r^{4})dr
\end{equation*}

Making use of the continuity at $r_{a}$, namely, $%
X_{m}-Y_{m}r_{a}^{2}=X_{m}^{(l)}-Y_{m}^{(l)}r_{a}^{2}$, we obtain

\begin{equation}
r_{b}=r_{a}(\frac{Y_{l}-Y_{m}+Y_{m}^{(l)}}{Y_{m}^{(l)}})^{1/5}  \label{eq12}
\end{equation}

\noindent and $X_{m}^{(l)}=Y_{m}^{(l)}r_{b}^{2}$. It is clear that, in order to have $%
u_{l}/r$ descending in the second domain,\ $Y_{m}^{(l)}>0$ is necessary to
be presumed. Together with the previously mentioned condition $Y_{l}\geq
Y_{m}$, $r_{b}$ is well defined from eq.(\ref{eq12}) and $r_{b}\geq r_{a}$
holds. Furthermore, once $X_{m}^{(l)}$ is known, $X_{m}$ can be known from
the continuity at $r_{a}$ as
\begin{equation*}
X_{m}=Y_{m}^{(l)}r_{b}^{2}+(Y_{m}-Y_{m}^{(l)})r_{a}^{2}  \label{eqx2}
\end{equation*}

Recall that $X_{m}>0$\ has been presumed. In order to recover this
presumption, the W-strengths should be so chosen to ensure

\begin{equation}
r_{b}/r_{a}>[(Y_{m}^{(l)}-Y_{m})/(Y_{m}^{(l)})]^{1/2}  \label{eq13}
\end{equation}

\noindent $u_{n}/r$ is distributed in three domains. In the first domain ($0,r_{a}$)
where all the wave functions are nonzero, it must have the form\ $%
u_{n}^{2}/r^{2}=X_{n}-Y_{n}r^{2}$. In the second domain ($r_{a},r_{b}$), $%
u_{n}^{2}/r^{2}=X_{n}^{(l)}-Y_{n}^{(l)}r^{2}$. While in the third domain ($%
r_{b},r_{c}\equiv \sqrt{2\varepsilon _{n}}$ ) only $u_{n}/r$ is nonzero and
appears as $u_{n}^{2}/r^{2}=\frac{1}{\alpha _{nn}}(\varepsilon _{n}-r^{2}/2)$%
. When $r=r_{c}$, $u_{n}/r$ arrives also at zero. Due to \textit{Lemma III},
$r_{c}$ is the outmost border for all the atoms. Making use of the
continuity at $r_{a}$ and $r_{b}$, $X_{n}$, $X_{n}^{(l)}$ and $\varepsilon
_{n}$ are related as $X_{n}=X_{n}^{(l)}+(Y_{n}-Y_{n}^{(l)})r_{a}^{2}$ and $%
X_{n}^{(l)}=\frac{1}{\alpha _{nn}}[\varepsilon _{n}-(\frac{1}{2}-\alpha
_{nn}Y_{n}^{(l)})r_{b}^{2}]$. Inserting these two relations into the
normalization $\int_{0}^{r_{c}}u_{n}^{2}dr=1$, we have

\begin{eqnarray}
\varepsilon _{n}=\frac{X_{l}}{2Y_{l}}[2\alpha
_{nn}(Y_{l}-Y_{n}-(Y_{l}-Y_{m})Y_{n}^{(l)}/Y_{m}^{(l)}) \nonumber \\
+1+(Y_{l}-Y_{m})/Y_{m}^{(l)}]^{2/5`}  \label{eq14}
\end{eqnarray}

\noindent Thus, $\varepsilon _{n}$ can be known when all the W-strengths are given.
After $\varepsilon _{n}$ is known, $X_{n}^{(1)}$ and $X_{n}$ can be known
from the continuity as shown above. Thus $u_{n}/r$ is obtained. Furthermore,
making use of eq.(\ref{eq6}) we have $\varepsilon _{l}=\Sigma _{l^{\prime
}}\alpha _{ll^{\prime }}X_{l^{\prime }}$. Thus, when all ${X_{l}}$\ are known,
all ${\varepsilon _{l}}$\ can also be known.

In the above form of $u_{n}/r$, $\varepsilon _{n}>r_{b}^{2}/2$ is required
to assure $\frac{u_{n}}{r}|_{r=r_{b}}>0$. Thus, from eqs.(\ref{eq14},\ref{eq12})
the W-strengths should be so chosen that
\begin{equation}
Y_{l}-Y_{n}>(Y_{l}-Y_{m})Y_{n}^{(l)}/Y_{m}^{(l)}  \label{eq15}
\end{equation}

\noindent is satisfied. In order to have $\frac{u_{n}}{r}|_{r=r_{a}}>0$ (i.e., $%
X_{n}^{(l)}>Y_{n}^{(l)}r_{a}^{2}$)
\begin{equation}
\varepsilon _{n}>\alpha _{nn}Y_{n}^{(l)}r_{a}^{2}+(1/2-\alpha
_{nn}Y_{n}^{(l)})r_{b}^{2}  \label{eq16}
\end{equation}

\noindent should be satisfied. In order to have $\frac{u_{n}}{r}|_{r=0}>0$ (i.e., $%
X_{n}>0$)

\begin{equation}
\varepsilon _{n}>\alpha _{nn}(Y_{n}^{(l)}-Y_{n})r_{a}^{2}+(1/2-\alpha
_{nn}Y_{n}^{(l)})r_{b}^{2}  \label{eq17}
\end{equation}

\noindent should be satisfied.

Thus, the miscible phase with all the three ${u_{l}/r}$ compactly surrounding
the center and with the ranges $r_{a}\leq r_{b}\leq r_{c}$\ will appear when
the W-strengths are so given that the conditions (i) $Y_{l}\geq Y_{m}$ and $%
Y_{l}\geq Y_{n}$. (ii) $Y_{m}^{(l)}>0$, and (iii) eqs.(\ref{eq13},\ref{eq15},%
\ref{eq16},\ref{eq17}) are satisfied. This specific miscible phase is
denoted as \{l,m,n\} to demonstrate that $u_{l}/r$\ has a narrowest
distribution and $u_{n}/r$\ has a broadest distribution.

\section{ Division of the parameter-space}

Obviously, the above inequalities together define a specific zone in the
multi-dimensional space of parameters. The borders of the zone are given by
the surfaces defined by the equalities arising from changing each of the
above inequality to equality. Note that the labels (l,m,n) can be any
permutation of (1,2,3). Therefore, there are six types of miscible states
and, correspondingly, six types of zones. Let the zones associated with
\{l,m,n\}=\{1,2,3\}, \{1,3,2\}, \{2,1,3\}, \{2,3,1\}, \{3,1,2\}, and
\{3,2,1\} be denoted as Zone I to Zone VI, respectively. The zone not
belonging to the above six is for the immiscible phase and is denoted as
Zone 0. Once the whole space has been divided into zones, one can select any
subset of parameters as variables while the others are fixed. This leads to
various types of phase-diagrams \ that demonstrate the variation of the
spatial configuration against the selected parameters. An example is given
in Fig.1. Note that it was found in 2-BEC that the g.s. might be asymmetric
when the interspecies interaction is sufficiently repulsive. This happens
when $\alpha _{lm}^{2}\geq \alpha _{ll}\alpha _{mm}$ (or $c_{ss^{\prime
}}\geq \sqrt{c_{s}c_{s^{\prime }}}$), and is expected to happen also in
3-BEC. Therefore, in Fig.1, $c_{ss^{\prime }}\leq \frac{3}{4}\sqrt{%
c_{s}c_{s^{\prime }}}$ are chosen to avoid the possible appearance of the
asymmetric configurations.

\begin{figure}[tbp]
\scalebox{0.8}{\includegraphics {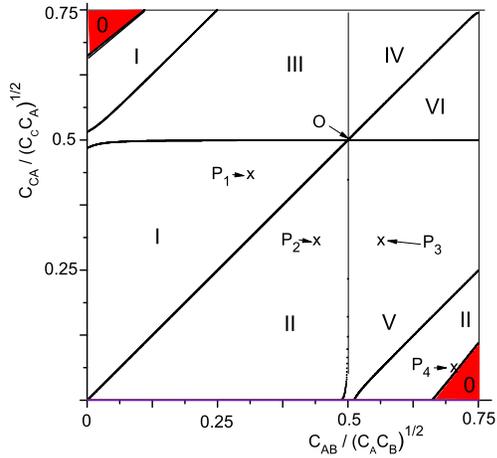}}
\caption{(Color online) Zones demonstrated in a 2-dimensional subspace expanded by $%
c_{AB}/(c_{A}c_{B})^{1/2}$ and $c_{CA}/(c_{C}c_{A})^{1/2}$. The other
parameters are $N_{A}=N_{B}=N_{C}=10^{4}$, $\protect\gamma _{A}=\protect%
\gamma _{B}=\protect\gamma _{C}=1$, $c_{BC}=10^{-3}\hbar \protect\omega
\protect\lambda ^{3}$, and $c_{A}=c_{B}=c_{C}=2c_{BC}$. The type of each zone is marked.
 The zone marked by 0 (in red) is for immiscible phase.}
\end{figure}

Due to the choice of the parameters, a number of symmetries are involved in
Fig.1.

(i) Let $c_{AB}/\sqrt{c_{A}c_{B}}$\ and $c_{CA}/\sqrt{c_{C}c_{A}}$\ be
denoted as $x$ and $y$. A reflection with respect to the axis $x=y$\ is
equivalent to the B- and C- atoms interchanging their names. Therefore, the
pattern is invariant against the reflection together with an interchange of
the indexes 2 and 3 (say, the zone \{1,2,3\} is changed to \{1,3,2\}), and
therefore Zone I is changed to II. Similarly, III$\leftrightarrow $V and IV$%
\leftrightarrow $VI.

(ii) When $x=1/2$, due to the specific choice of the parameters, $\alpha
_{12}=\alpha _{21}=\alpha _{23}=\alpha _{32}$. In this case the symmetry
inherent in the CGP assures $Y_{1}=Y_{3}$, and $u_{1}=u_{3}$. Thus, for the two labels
 {1,3,2} and {3,1,2} (they are related to each other by interchanging 1 and
3), the g.s. can be denoted by either one of them at the axis $x=1/2$.
 Accordingly, once Zone II appears in one side of the axis, Zone V will
also appear in the other side as its partner. Similarly, III and IV are
partners. In general, the
axis $x=1/2$ is replaced by a surface $Y_{1}=Y_{3}$ in the parameter-space.
 On the surface $u_{1}$ and $u_{3}$ overlap.

(iii) Similarly, when the axis $y=1/2$ be the common border of two
neighboring zones, the labels for these two zones are related to each
other by an interchange of 1 and 2 (say, \{1,2,3\} and \{2,1,3\}). Thus, I
and III are partners. V and VI also. As before, one can prove that, $u_{1}$
and $u_{2}$ overlap at the horizontal line $y=1/2$ (or, in general, on the
surface $Y_{1}=Y_{2}$).

(iv) The point $O$ is the intersection of the $x=1/2$ and $y=1/2$ axes (in
general, the intersection of the two surfaces $Y_{1}=Y_{2}$ and $Y_{2}=Y_{3}$%
), where all the $\alpha _{ll^{\prime }}$ are equal, and the three wave
functions ${u_{l}}$ overlap. Accordingly, in the neighborhood of $O$ all the
six types \{l,m,n\} have an equal probability to appear as shown in the
figure.

\begin{figure}[tbp]
\scalebox{0.8}{\includegraphics {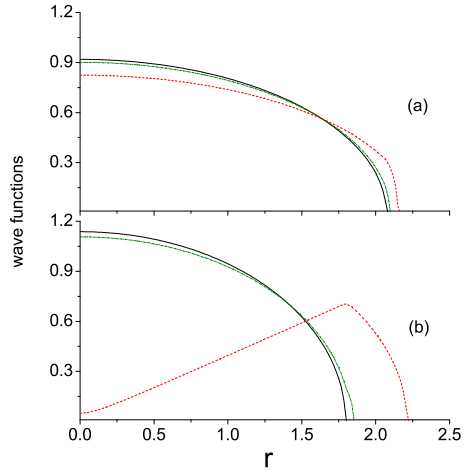}}
\caption{(Color online) $u_{1}/r$\ (solid), $u_{2}/r$\ (dash), and $u_{3}/r$%
 (dash-dot-dot) are plotted against $r$. The unit of $r$\ is $\protect%
\lambda \equiv \protect\sqrt{\hbar /(m\protect\omega )}$. (a) is associated
with the point P$_{2}$ marked in Fig.1, (b) is associated with P$_{4}$.}
\end{figure}

Since the solutions have been obtained in an analytical way, it is straight
forward to plot the wave functions. Examples are shown in Fig.2. In 2a the
wave functions are associated with the point P$_{2}$ marked by a cross in Fig.1, where
the g.s. is in the \{1,3,2\} phase. The pattern associated with P$_{1}$ is
identical with that of 2a but $u_{2}$ and $u_{3}$ interchange. The pattern
associated with P$_{3}$ is close to 2a (not exactly the same) but $u_{1}$
and $u_{3}$ interchange. All the three points are not far away from the
point O. Therefore the three wave functions are not remarkably different
from each other. Otherwise, they might be very different. 2b is associated
with P$_{4}$, where the g.s. is also in the \{1,3,2\} phase. However, due to
P$_{4}$ is very close to the zone of immiscible phase, the B-atoms tend to
leave completely from the center and tend to form a shell as shown by the
dash curve.

It was found that in the neighborhood of the border separating the miscible
and immiscible phases, the configuration is very sensitive to the variation
of parameters. For an example, $P_{4}$\ (marked in Fig.1) has $x=0.71$, and
accordingly $(u_{2}/r)_{r=0}=0.048$ (it implies that the B-atoms are very few
 at the center). When $P_{4}$\ shifts a little away from
the border so that $x$ becomes $0.70$, $(u_{2}/r)_{r=0}$ becomes 0.207. Thus, the
neighborhood of the above border is a region of sensitivity. In this region
a tiny change in the parameters) might cause an explicit change in the configuration. The existence
of regions of sensitivity in the parameter-space is a notable phenomenon.

\begin{figure}[tbp]
\scalebox{0.8}{\includegraphics {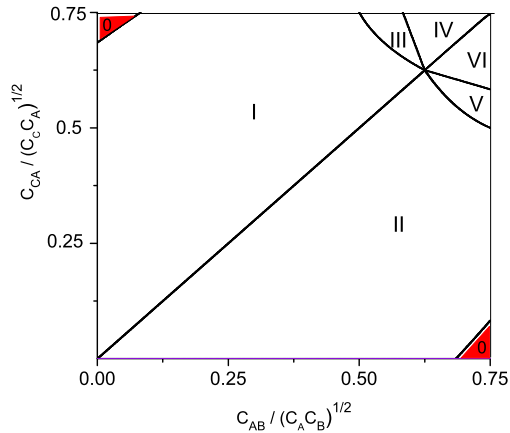}}
\caption{(Color online) The same as Fig.1 but with $N_{B}=N_{C}=15000$ while
$N_{A}$ remains to be 10000.}
\end{figure}

One more example is given in Fig.3 to demonstrate the effect of particle
numbers. In this figure the number of A-atoms is smaller. The symmetry with respect to the $x=y$ axis (i.e.,
an interchange of 2 and 3) remains, while the other symmetries appear no more.
Note that the Zone I and II are dominant in Fig.3 implying that the A-atoms are closer
to the center. Thus, when all the three ${c_{s}}$ are close to each other and
the three ${\gamma _{s}}$ also, the kind of atoms with least particle number
will be closer to the center.

\section{ Final remarks}

We have succeeded to derive the analytical solutions of the CGP for the
3-BEC under the TFA . Thereby the parameter-space has
been divided into zones each supports a specific configuration. Based on the division, various types
of phase diagrams can be plotted, and the
variation of the spatial configurations against the parameters can be visualized.
 From the experience of 2-BEC, when the particle numbers are large and when both kinds
of atoms are distributed surrounding the center (i.e., ${u_{l}}$ are nonzero when $r=0$), the wave functions
obtained under TFA and beyond TFA overlap nearly completely (refer to Fig.1a and 1b of
\cite{polo15}, where a detailed discussion on the accuracy of the TFA is made). Since this paper concerns only this case
, the TFA is believed to be applicable. Nonetheless, this remains to be further clarified.

Obviously, this paper is far from a complete description of the 3-BEC.
Note that, when the inter- and intra- species interactions are close in strengths
or the former is stronger than the latter, Symmetric immiscible states and
asymmetric states may emerge. The
details and the classification of these states remain to be studied.

The variety of the spatial configurations of 3-BEC implies that rich physics
is involved. In particular, just as in 2-BEC, regions of sensitivity have
been found. When a realistic parameter falls in a region of sensitivity, it
can be more accurately determined. Obviously, these regions of 2-BEC and
3-BEC are different. Thus, in addition to the 2-BEC, the 3-BEC will be
helpful in the determination of parameters. Recall that the BEC are a
valuable tool because they are tunable. One can consider that the addition
of the third kind of atoms into a 2-BEC is an one more way to tune the system.

Incidentally, the above analytical approach is quite general and can be
generalized to deal with the condensates with more than three species.

\begin{acknowledgments}

Supported by the National Natural Science Foundation of China under Grants
No.11372122, 11274393, 11574404, and 11275279; the Open Project Program of
State Key Laboratory of Theoretical Physics, Institute of Theoretical
Physics, Chinese Academy of Sciences, China(No.Y4KF201CJ1) ; and the National Basic Research
Program of China (2013CB933601).

\end{acknowledgments}

\bibliography{basename of .bib file}

\end{document}